\documentclass[11pt,fleqn]{article}

\usepackage{latexsym}
\usepackage{amssymb,amsfonts}
\usepackage[dvips]{graphicx}
\usepackage{cite}
\usepackage{marvosym}
\usepackage{euscript} 
\usepackage{bm}        
\usepackage{amsmath}   

\catcode`\@=11
\@addtoreset{equation}{section}
\def\theequation{\arabic{section}.\arabic{equation}}

\def\appendix{\renewcommand{\thesection}{\Alph{section}}\setcounter{section}{0}
              \renewcommand{\theequation}
            {\mbox{\Alph{section}.\arabic{equation}}}\setcounter{equation}{0}}

\def\babs{\hrule\par\begin{description}\item{Abstract: }\it} 
\def\eabs{\par\end{description}\hrule\par\medskip\rm}

\newcommand{\s}[1]{\section{#1}}

\def\eeq{\end{eqnarray}}      
\def\beq{\begin{eqnarray}}
               
\def\at{\left(}

\def\ct{\right)}

\def\R{{\hbox{{\rm I}\kern-.2em\hbox{\rm R}}}}   
\def\H{{\hbox{{\rm I}\kern-.2em\hbox{\rm H}}}}   
\def\N{{\hbox{{\rm I}\kern-.2em\hbox{\rm N}}}}   
\def\C{{\ \hbox{{\rm I}\kern-.6em\hbox{\bf C}}}} 
\def\Z{{\hbox{{\rm Z}\kern-.4em\hbox{\rm Z}}}}   

\def\La{\Lambda}

\begin{document}

\title{$\La$ may not be vacuum energy, after all}
\author{Luciano Vanzo\thanks{email: luciano.vanzo@unitn.it}    \\ 
Dipartimento di Fisica, Universit\`a di Trento\\ 
and TIFPA, INFN Center, Via Sommarive 14\\
 38123 Povo, Trento, Italia\\
\textit{Essay written for the Gravity Research Foundation 2014 Awards for Essays on Gravitation}}

\date{March 31, 2014}

\maketitle


\begin{abstract}
We suggest the possibility that the mysterious dark energy component driving the acceleration of the Universe is the leading term, in the de Sitter temperature, of the free energy density of space-time seen as a quantum gravity coherent state of the gravitational field. The corresponding field theory classically has positive pressure, and can be considered as living on the Hubble horizon, or, alternatively, within the non compact part of the Robertson-Walker metric, both manifolds being characterized by the same scale and degrees of freedom. The equation of state is then recovered via the conformal anomaly. No such interpretation seems to be available for negative $\Lambda$.
\end{abstract}

\s{Introduction}

Many papers have been written on the vast subject of the vacuum energy but,  still, a convincing resolution is lacking. The vacuum energy in quantum field theory, naively computed by summing zero point energies, is much more larger than observed, and a fantastic fine tuning seems necessary to adjust its value to the physical reality\cite{Weinberg:1988cp}.  It has also the enduring quality of being an indestructible constant, so that once it is there, will be for ever. 
Of course, in saying this, we assume that we are really observing the vacuum. But may be not. It is conceivable  that the presence of a cosmological constant has to do with a vacuum-like state of the sort encountered in inflationary theories, that possesses a certain amount of energy density but no elementary particles whatsoever. In the case of inflation, no one knows how such a state formed in the early Universe, but a classical coherent field would be a natural effective picture. It could also have a temperature, despite the absence of elementary particles, in the form of a parameter encoded within the vacuum correlations: that is, the power spectrum of the vacuum correlations has a nearly Planckian distribution. We all recall the Rindler vacuum, that has a temperature but no elementary particles. Instead, it has a noise, whose thermal nature can be seen by an accelerated observer, but  inertial observers cannot detect  (see the comprehensive reviews\cite{Takagi:1986kn,Crispino:2007eb}). It can also be interpreted as the temperature characterizing the Minkowski vacuum seen as a KMS state, relatively to a different time evolution\cite{Bisognano:1976za,Sewell:1982zz}. 
  
In the following sections, by taking inspiration from a relation between de Sitter gravity and Liouville theory that we discussed long ago\cite{Klemm:2002ir,Klemm:2004mb}, and from Ginsparg \& Perry suggestion\cite{Ginsparg:1982rs} that ``de Sitter space might be expected to behave in some respects like flat space at finite temperature'', we want to explore the idea that the state of the Universe in which the dark energy effectively drives the expansion, could be such a vacuum without gravitons, holding nonetheless a kind of energy density that can even be degraded or exchanged with other forms of matter. Along the way, this certainly soften the uncomfortable ``lasting for ever'' quality mentioned above. 

However, the interpretation of this energy as a kind of thermal energy entails an apparent contradiction with cosmology, because then the thermodynamic pressure is positive. The issue is resolved by making use of the conformal anomaly. As a bonus, if the expansion flattens the horizon then the anomaly is driven to zero, saving the Universe from what looks like a fatal, eternally accelerated expansion.

\s{Vacuum energy, thermal energy and the conformal anomaly}

If the dark energy component of the universe is controlled by a
positive cosmological constant, there will be an event horizon
whose size is of order $L=\sqrt{3/\La}$. As it is well known, in this case the energy density and pressure are given by, respectively,
\[
\rho_{\La}=\frac{\La}{8\pi G}, \qquad p=-\rho_{\La}
\]
Now the event horizon is not really in equilibrium with the
cosmological background radiation, since quantum effects
imply a far lower characteristic temperature\cite{Gibbons:1977mu},  
\[
T=\frac{1}{2\pi L}=\frac{H_0}{2\pi}=\frac{\sqrt{\La/3}}{2\pi}\simeq 10^{-42}\;\rm{Gev}\,,
\]
the last equality holding at the present times. This is the famous de Sitter temperature. More generally, in a Universe which is only asymptotically dominated by the vacuum energy, we have the following scale factor\footnote{This comes out from an elementary integration of the Friedmann equations in a cosmology with flat spatial sections and negligible radiation.} (normalized to one at the present time $t_{0}$)
\beq\label{qdsa}
a(t)=\left(\frac{\Omega_{m}}{\Omega_{\Lambda}}\right)^{1/3}\sinh^{2/3}\left(\frac{3}{2}\sqrt{\Omega_{\Lambda}}H_{0}t\right), 
\eeq
where $H_{0}$ is the present value of the Hubble constant.  Looking at the asymptotic form of the scale factor, we see that it corresponds to a de Sitter expansion with a bit different effective temperature
\beq\label{qdsT}
T=\sqrt{\Omega_{\Lambda}}\,\frac{H_{0}}{2\pi}=\frac{1}{2\pi L}
\eeq
where, as usual, $\Omega_{\Lambda}$ is the (present)  fraction of dark energy over the critical density and the second equality defines $L$. One can easily see that $\sqrt{\Omega_{\Lambda}}H_{0}$ is the asymptotic value of the Hubble parameter.  Actually, this result holds true for any homogeneous and isotropic cosmological metric with a trapping horizon, in the sense given to it by Hayward\cite{Hayward:1993wb,Hayward:1997jp}. If we take as the temperature parameter that given by the corresponding surface gravity, then one finds\cite{DiCriscienzo}
\beq\label{temp}
T=\frac{1}{2\pi}\left(H^2+\frac{1}{2}\dot{H}+\frac{\kappa}{2a^2(t)}\right)\left(H^2+\frac{\kappa}{a^2(t)}\right)^{-1/2}
\eeq
If the Universe is asymptotically dominated by a constant vacuum energy, then the curvature term and $\dot{H}$ become negligible, and so $T\to H_{\rm{asym}}/2\pi=\sqrt{\Omega_{\Lambda}}H_{0}/2\pi$. As a partial justification for taking seriously the above expression of the temperature, we remark that there is a rate for particle creation near a trapping horizon that has the form of a Boltzmann tail, $\Gamma\simeq\exp(-E/T)$, with $T$  given by Eq.~\eqref{temp}\cite{DiCriscienzo}.

This prompt us to rewrite the vacuum energy as thermal field energy. Noting that the critical density is $3 H_{0}^{2}M_{p}^{2}/8\pi$, from Eq.~\eqref{qdsT} we get ($M^{2}_{p}=G^{-1}$)
\beq\label{then}
\rho=\at\frac{2\pi}{H_{0}}\ct^{2}\at\frac{3H_{0}^{2}}{8\pi G}\ct\,T^{2}=\frac{3\pi}{2G}\,T^2=\frac{3}{2}\pi M_p^2T^2
\eeq
suggesting a comparison with two-dimensional quantum field theory. This is our first, very simple, observation. Actually, the free energy of thermal gravitons has a leading term of order $T^4$, plus sub-leading terms of order $T^2$, but here we refuse the interpretation in terms of real gravitons. Rather, we assume that $\rho$ is itself the leading term of the energy density of space-time, in the asymptotic limit. From this derives our insistence on $2D$ field theory. That this is possible even in $4$ dimensions is related of course to the presence of the Planck scale $M_{p}^{2}$, that supplies the correct dimensional analysis. 

Now, if considered on a compact space of size $L$, such a theory has an energy density
whose leading term is
\beq\label{cch}
\rho=\frac{\pi c}{6}\,T^2
\eeq
where $c$ is known as the (unknown) central charge. We will consider this as the right formula for non compact spaces too, by using the usual argument of the thermodynamics limit.

Now massless radiation in two dimensions apparently has pressure $p=\rho$, so the comparison
blatantly contradicts cosmology which needs $p=-\rho$, under our
hypothesis that the dark energy is exactly constant, or asymptotically constant. How can a  field theory have positive energy but negative pressure? A possible answer, in two-dimensions, is the conformal anomaly.


In fact, the reason of $p=\rho$ in a classical $2D$ field theory is rooted in the
fact that the trace of the energy momentum tensor of such a theory classically vanishes
\[
T^{\mu}_{\;\mu}=-\rho+p=0
\]
But of course this must be corrected with the conformal anomaly, that  we interpret here as a quantum breaking of the classical relation $p=\rho$. For a theory with a central charge $c$, this is (see, for example, Ref.~\cite{polchinski2005})
\beq\label{qch}
T^{\mu}_{\;\mu}=-\rho+p=-\frac{cR}{24\pi}
\eeq
From Eqs.~\eqref{then} and \eqref{cch} we read off the central charge, $c=9M_p^2$. If the curvature radius of our two-dimensional space is $L$, then
\beq\label{curv}
R=2L^{-2}=8\pi^2T^2
\eeq
and the anomaly equation becomes
\beq\label{result}
p=\rho-\frac{9M_{p}^{2}}{24\pi}\,8\pi^2T^2=-\frac{3}{2}\,\pi M_{p}^{2}T^2=-\rho
\eeq
in agreement with accelerated cosmology. Note that in a general conformal field theory there is not any fundamental reason for this result.  
To repeat, the  thermal interpretation of the energy density provides a central charge; once combined with the anomaly this gives the equation of state of the dark energy component. Sub-leading corrections to the dominant quadratic term would slightly change the exact relation $p+\rho=0$, hopefully in the right direction. So our tentative conclusion is that an excited, two-dimensional, thermal state  associated to cosmology, can have a negative pressure of the right magnitude to act as a form of dark energy. 
 
\s{The place of the two-dimensional field theory, and the conclusions}

This title summarizes the weak point of this business. After the string theory lesson, one can imagine a variety of ways in which a correspondence between four-dimensional physics and a quantum two-dimensional world can be established. There seems to be only few possibilities:  some unitary field theory with a built-in mass scale (the Planck mass giving the needed central charge), living on the horizon two-sphere or, alternatively, on the $2$-section of the cosmological field described by the metric tensor and curvature scalar,
\beq\label{frw2}
ds^{2}=-dt^{2}+a(t)^{2}\frac{dr^{2}}{1-\kappa r^{2}}, \quad R=\frac{2\ddot{a}}{a},
\eeq
respectively.  Both manifolds are characterized by the same degrees of freedom, the dynamical scale factor $a(t)$, and both have curvature radius of order  $L^{-1}=\sqrt{\Omega_{\Lambda}}\,H_{0}$. Moreover, bulk solutions with cosmological event horizons, like those considered here, correspond to microscopic states of quantum two-dimensional Liouville theory\cite{Klemm:2002ir,Klemm:2004mb}, whereas those without horizons correspond to macroscopic (normalizable) states, with no underlying statistical description.

The horizon expanding sphere looks like as the more natural choice from the hamiltonian viewpoint, since the hamiltonian of general relativity for the interior region has a boundary observable (the York energy), which can be computed and is the total dark energy enclosed within.  

The coupling of the dark energy to $a(t)$, provided by the conformal anomaly, offers a possible way out to the frightening,  eternally accelerated expansion. If the expansion could  flatten the horizon the anomaly would be driven to zero. In this way, the dark energy component would be gently converted into a more conventional form of thermal energy. \\

The author would like to acknowledge fruitful conversation with S. Zerbini.


\end{document}